\documentclass[a4paper,11pt]{article}
\usepackage{pos}

\title{A brief survey of low energy supersymmetry under current experiments}
\ShortTitle{Low energy SUSY under current experiments}

\author*{Jin Min Yang}
\author[]{Pengxuan Zhu}
\author[]{Rui Zhu}
\affiliation[ ]{CAS Key Laboratory of Theoretical Physics, Institute of Theoretical Physics, \\
Chinese Academy of Sciences, Beijing 100190, P.R.China}

\affiliation[ ]{School of Physical Sciences, University of Chinese Academy of Sciences,  \\
Beijing 100049, P.R.China}

\emailAdd{jmyang@itp.ac.cn}
\emailAdd{zhupx99@icloud.com}
\emailAdd{zhurui@itp.ac.cn}

\abstract{This is a brief overview on the low energy supersymmetry 
in light of current experiments including the LHC searches, the dark matter (DM) detections, 
the muon $g-2$ and the CDF II measurement of the $W$-boson mass. 
We focus on the minimal framework of  supersymmetry, namely the minimal supersymmetric model (MSSM), 
and obtain the following conclusions:
(i) The MSSM can survive all current experiments, albeit suffering from the little hierarchy problem 
due to the heavy stops pushed up by the LHC searches; (ii) The DM relic density can be readily achieved by 
the thermal freeze-out of the lightest neutralino and the null results of DM direct detections are typically 
driving the parameter space to the bino-like lightest neutralino region; (iii) The muon $g-2$ anomaly 
reported by FNAL and BNL can be explained at $2\sigma$ level, which indicates light sleptons and 
electroweakinos possibly accessible at the HL-LHC;  
(iv)  The CDF II measurement of the $W$-boson mass can be marginally explained, but requires light stops 
near TeV which may soon be covered by the LHC searches.  } 
 
\FullConference{%
  The Tenth Annual Conference on Large Hadron Collider Physics - LHCP2022\\
  16-20 May 2022\\
  online
}

\begin{document}
\maketitle
{\em Introduction}:
The discovery of Higgs boson structurally completed the Standard Model (SM). 
Yet, it is widely believed that the SM is not the ultimate theory because it has many theoretical problems. 
So the main task of high energy physics has become the test of the Higgs property and the searches for new physics beyond the SM. 
So far new physics has been intensively explored. 
Although the ``zero'' results from the LHC experiments and the dark matter direct detections have not shaken our belief in the existence of new physics, 
{\em physics thrives on crisis and we all recall the great progress made while finding a way out of the various crises of the past}~\cite{weinberg}. 
In order to move forward, the crises from experiments are needed. 
Recently, there have appeared two harbingers {\em indirectly} indicating the existence of new physics:
the muon anomalous magnetic moment reported by the Fermilab result~\cite{Muong-2:2021ojo} (combined the BNL data) and the W-boson mass reported by CDF II \cite{CDF:2022hxs} are above the SM theoretical predictions by  $4.2\sigma$ and $7\sigma$, respectively.  
Unlike the non-zero neutrino masses which usually imply superheavy right-handed neutrinos, these two anomalies may suggest the new physics at TeV-scale, and thus have injected strong excitement into the 
field of high energy physics.  
Among various new physics theories, the low energy supersymmetry (SUSY) has been a mainstream due to its graceful mathematical structure and theoretical framework. 
Indeed, {\it SUSY is part of a larger vision of physics, not just a technical solution} \cite{witten}. 
Some SUSY parameter space has been excluded by the LHC and dark matter direct detection experiments. 
The on-going LHC and the forthcoming HL-LHC will keep searching for super particles, and hence the 
phenomenological studies of low energy SUSY remain a hot topic in high energy physics. 
In this note, we focus on the minimal framework of SUSY, namely the minimal supersymmetric 
model (MSSM) \cite{Haber:1984rc},
to give a brief overview (for recent long reviews see, e.g., \cite{Wang:2022rfd,Baer:2020kwz}) 
on the current status of low energy SUSY. 
\vspace{0.01cm}

{\em A brief description of MSSM}:
As well known, SUSY algebra extends the Poincare group by anti-commutators, which is the only graded Lie algebra of the S-matrix symmetries consistent with the relativistic quantum field theory. 
In other word, SUSY is an {\it extension of special relativity to include fermionic symmetries} \cite{witten-bj}.
Thus SUSY predicts that every SM particle has a super partner. 
Among various SUSY models, the MSSM is the most economic one with a minimal particle content. 
But even this minimal framework predicts five Higgs bosons, in comparison with the SM which predicts only one Higgs boson. 
Among the super particles, the super partners of the top quark (called stops) are rather unique, which usually are the lightest colored sparticles and directly related to the naturalness. 
The charginos and neutralinos (called electroweakinos) also merit special attention because they are likely to be lighter than the colored sparticles. 
For example, the higgsino-like sparticles in natural SUSY (see, e.g., \cite{Tata:2020afe}) typically have light masses of 100-300 GeV. 
In addition, the lightest neutralino is usually assumed to be the lightest sparticle (LSP) and serve as the cold dark matter candidate. 
Since the experiments have ruled out the prediction of sparticles as light as their SM partners, SUSY must be spontaneously broken in some hidden sector and then the breaking effects are mediated via gauge interactions (GMSB), via anomaly (AMSB) or via gravitational interaction (SUGRA) to the visible sector. 
Without assuming a certain breaking or mediation mechanism, the MSSM has over 100 free parameters, which are mainly from the soft breaking terms. 
Once the boundary conditions are imposed by a certain mechanism, the model such as the constrained MSSM (CMSSM) has only a few free parameters and thus has a strong predictive power. 
\vspace{0.1cm}
      
{\em MSSM confronted with LHC searches}:
At the LHC the signature of MSSM with R-parity is the multi-jet and/or multi-lepton plus missing energy (carried out by the LSP ) final states from the sparticle pair productions with cascade decay chains. 
The colored sparticles (squarks and gluinos) could be copiously produced by strong interaction, while the uncolored sparticles (electroweakinos and sleptons) with electroweak interaction have relatively small production rates.
Thus the non-observation of sparticle production at the LHC has pushed the colored sparticles above about 2 TeV \cite{CMS:2020cur} (except stops discussed below), while the bounds on uncolored sparticles are a few hundreds or even around one hundred GeV \cite{ATLAS:2020pgy,ATLAS:2019lff,ATLAS:2019wgx,GAMBIT:2018gjo}.
Of particular notes are stops which are closely related to naturalness and the lightest neutralino which usually serves as the dark matter condidate. 
For the stop (refer to the lighter mass eigenstate $\tilde t_1$), from the Run-II results we see that the bounds on the stop mass depend on the LSP mass \cite{ATLAS:2017eoo,CMS:2019ysk}. 
For a LSP lighter than about 350 GeV, the stop mass is pushed above 1.2 TeV.
But for the tough regions where the stop mass is close to the LSP mass (called the LSP-stop compressed spectrum) or close to the LSP mass plus the top mass (called the stealth stop), a stop as light as 600 GeV is still allowed. 
In the former case, the stop merely appears as missing energy and their pair production can only be probed by an associated energetic jet in final state or by other correlated processes \cite{Duan:2019ykd}.
In the latter case the advanced machine learning techniques were shown to improve the signal sensitivity \cite{Abdughani:2018wrw}.    
For the lightest neutralino, if it is higgsino-like as in natural SUSY, then the LSP $\tilde\chi^0_{1}$ and the NLSP $\tilde\chi^0_{2}$ as well as the charginos $\tilde\chi^{\pm}_1$ are all nearly degenerate around the mass value $\mu$. Their productions at the LHC just give a signal of  
missing energy plus very soft particles, so their searches are rather challenging, with the current LHC bound merely being around 100 GeV  \cite{Han:2013usa, Lv:2022pme, Buanes:2022wgm}.
The most remarkable achievement of LHC is the discovery of a 125 GeV Higgs boson. This is a strong support of low energy SUSY because the MSSM predicts a CP-even neutral Higgs boson lighter than $Z$-boson mass at tree level and upper bounded by about 135 GeV at loop level. 
Since a 125 GeV Higgs boson requires sizable quantum effects from relatively heavy stops (say above TeV for a moderate trilinear coupling $A_t$), the non-observation of colored sparticles like stops at the LHC is just consistent with the requirement of the 125 GeV Higgs boson.            
Since the stop masses are significantly above the weak scale of 100 GeV, the logarithmic divergence in the loops of the Higgs mass, which is proportional to the stop masses, becomes sizable, and thus the MSSM has the little hierarchy problem with a fine-tuning extent $\Delta_{\rm BG}$ at the per mille level \cite{Han:2016gvr} (the fine-tuning extent can be lower \cite{Baer:2020kwz} in term of another measure $\Delta_{\rm EW}$ \cite{Baer:2012up,Baer:2013gva}).  
\vspace{0.1cm}
       
{\em MSSM confronted with dark matter:}
In the MSSM, the lightest neutralino is usually assumed to be the LSP and thus serve as the cold DM candidate. 
It is a perfect weakly interacting massive particle (WIMP) candidate and can naturally provide a thermal relic abundance in agreement with the Planck measurement. 
However, such a neutralino WIMP typically scatters off the nucleon with weak interaction, giving a cross section accessible at the current DM direct detection experiments.
So the null search results from the DM direct detections typically pushed the LSP with a moderate mass into the bino-like region (the higgsino-like LSP with a mass around 1 TeV \cite{marco,hisano} and the wino-like LSP with a mass around 2.9 TeV \cite{hisano,wino-lsp2,wino-lsp3} are also viable and can provide correct relic abundance). 
For the bino-like LSP, the main concern is DM over-abundance. To speed up its annihilation rate in the early universe, the most popular way is wino-bino, slepton-bino or stop-bino coannihilation. 
If we do not require the LSP to provide a sufficient DM abundance, the higgsino-like LSP with a mass of 100-300 GeV in natural SUSY can still be a cold DM component since its scattering rate with the nucleon is much suppressed by its low relic density. 
Once we try to use bino-higgsino admixture to provide a sufficient DM abundance in natural SUSY, the LSP will scatter with the nucleon efficiently and can hardly satisfy the current direct detection limits \cite{Abdughani:2017dqs}. 
Overall, we can say that so far the DM relic density can be achieved by the thermal freeze-out of the neutralino LSP and the null results of DM direct detections typically pushed the parameter space into a relatively narrow bino-like LSP region.                 
\vspace{0.1cm}

{\em MSSM confronted with muon $g-2$ and W-mass:}
Both the muon $g-2$ and W-mass anomalies can be explained at 2$\sigma$ level in the MSSM.
For the explanation of muon $g-2$, a typical scenario is light bino-like LSP (coannihilated by wino to give correct DM relic density) plus not-too-heavy sleptons \cite{Abdughani:2019wai,Cox:2018qyi,Athron:2021iuf,Wang:2021bcx}. 
In this case, the light winos can be pair produced and decay to the LSP to give soft leptons plus missing energy, whose parameter space is found to be largely accessible at the HL-LHC \cite{Abdughani:2019wai}. 
If both the muon $g-2$ and electron $g-2$ measured by the Berkeley experiment are to be explained at $2\sigma$ level, a very specific parameter space is required \cite{Li:2021koa,Li:2022zap}, in which the thermal freeze-out of the bino-like LSP gives an over-abundance for the dark matter (in this case the LSP must late decay to a lighter superWIMP dark matter candidate).
For the explanation of W-mass anomaly measured by the CDF II \cite{Yang:2022gvz,Bagnaschi:2022qhb}, the loop contributions from stops must be sizable \cite{Yang:2022gvz,Heinemeyer:2013dia} and hence require the stop ($\tilde t_1$) around 1 TeV, which should be accessible at the LHC or HL-LHC.  
If the muon $g-2$ is to be explained simultaneously with the W-mass, the LSP cannot be heavy and such a stop around 1 TeV is quite near the current LHC exclusion bound. 
Otherwise, the LSP can be heavier and the LHC bound on stop mass can get weaker.
It should be noted that the MSSM with the usual boundary conditions at some high energy scales, such as the CMSSM, mSUGRA, GMSB and AMSB, cannot explain the muon $g-2$ or W-mass anomaly because the stops are heavy, as required by the 125 GeV Higgs mass, and the sleptons are also heavy due to the boundary conditions \cite{Wang:2021bcx,gm2-gutsusy-2,gm2-gutsusy-3} (extensions such as the gluino-SUGRA \cite{gm2-SUGRA-ext-1,gm2-SUGRA-ext-2,gm2-SUGRA-ext-3,gm2-SUGRA-ext-4}, the GMSB with Higgs-messenger 
coupling \cite{GMSB-yukawa-higgs-1,GMSB-yukawa-higgs-2} and the deflected AMSB \cite{gm2-SUGRA-ext-5,gm2-SUGRA-ext-6},  
can weaken the mass bound on the stops or sleptons). 

For the future exploration of low energy SUSY, in addition to the LHC and HL-LHC, 
a future Higgs factory can also indirectly probe the allowed parameter space via precision measurements of the Higgs properties \cite{Athron:2022uzz,Li:2020glc,Li:2022bbq}. Note that we restricted our discussions to the minimal framework of SUSY. For other well motivated extensions of the MSSM, such as the NMSSM \cite{Ellwanger:2009dp}, the pressure of the 125 GeV Higgs mass on the stops can be alleviated \cite{Cao:2012fz}, the explanation of muon $g-2$ can be readily achieved \cite{gm2-nmssm-1,gm2-nmssm-2,gm2-nmssm-3,Wang:2021lwi,Cao:2022htd,Tang:2022pxh,Cao:2018rix}, and the tight limits from the DM experiments can be relaxed a lot \cite{Cao:2019qng}.      
   
{\em Conclusion:} We conclude that the MSSM can survive all current experiments, albeit suffering from the little hierarchy problem due to the heavy stops pushed up by the LHC searches. 
The dark matter relic density can be achieved by the thermal freeze-out of the lightest neutralino and the dark matter direct detection limits can be satisfied by the bino-like neutralino. 
The muon $g-2$ anomaly can be explained at $2\sigma$ level, which indicates light sleptons and electroweakinos possibly accessible at the HL-LHC.  
The CDF II measurement of the $W$-mass can be marginally explained, but requires light stops near TeV which may soon be covered by the LHC searches. 
\vspace{0.1cm}

{\em Acknowledgments:} This work was supported by the National Natural Science Foundation of China 
(NNSFC) under grant Nos. 11821505 and 12075300.

\end{document}